\documentclass[preprint,superscriptaddress,noshowpacs,longbibliography]{revtex4-1}
\usepackage{amssymb}
\usepackage{amsmath}
\usepackage{amsfonts}
\usepackage{bm}
\usepackage{graphicx}
\usepackage[dvipsnames]{xcolor}
\usepackage{calc} 
\usepackage{epsfig}
\usepackage{epstopdf}
\begin{document}

\title{Detecting Topological Phases of Microwave Photons in a Circuit Quantum Electrodynamics Lattice}

\author{Yan-Pu Wang}
\affiliation{School of Physics and Wuhan National Laboratory for Optoelectronics, Huazhong University of Science and Technology, Wuhan, 430074, China}

\author{Wan-Li Yang}
\affiliation{State Key Laboratory of Magnetic Resonance and Atomic and Molecular Physics, Wuhan Institute of Physics and Mathematics, Chinese Academy of Sciences, Wuhan 430071, China}

\author{Yong Hu}
\email{huyong@mail.hust.edu.cn}
\affiliation{School of Physics and Wuhan National Laboratory for Optoelectronics, Huazhong University of Science and Technology, Wuhan, 430074, China}

\author{Zheng-Yuan Xue}
\email{zyxue@scnu.edu.cn}
\affiliation{Guangdong Provincial Key Laboratory of Quantum Engineering and Quantum Materials, School of Physics and Telecommunication Engineering, South China Normal University, Guangzhou 510006, China}

\author{Ying Wu}
\affiliation{School of Physics and Wuhan National Laboratory for Optoelectronics, Huazhong University of Science and Technology, Wuhan, 430074, China}

\begin{abstract}

Topology is an important degree of freedom in characterizing electronic systems. Recently, it also brings new theoretical frontiers and many potential applications in photonics. However, the verification of the topological nature is highly nontrivial in photonic systems as there is no direct analog of quantized Hall conductance for bosonic photons. Here we propose a scheme of investigating topological photonics in superconducting quantum circuits by a simple parametric coupling method, the flexibility of which can lead to the effective \textit{in situ} tunable artificial gauge field for photons on a square lattice. We further study the detection of the topological phases of the photons. Our idea employs the exotic properties of the edge state modes which result in novel steady states of the lattice under the driving-dissipation competition. Through the pumping and the photon-number measurements of merely few sites, not only the spatial and the spectral characters, but also the momentums and even the integer topological quantum numbers with arbitrary values of the edge state modes can be directly probed, which reveal unambiguously the topological nature of photons on the lattice.

\end{abstract}


\maketitle

\section*{Introduction}
\label{Sec Intro}

Charged particles in two dimension exhibit integer quantum Hall effect (IQHE) when exposed to a perpendicular magnetic field \cite{KlitzingPRL1980}, characterized by the quantized transverse conductances in transport experiments. This novel effect can be explained by the integer topological Chern numbers describing the global behavior of the energy bands \cite{TKNN1982PRL,BernevigBook}. Such topological insulating IQHE phase is robust against disorder and defects because the band topology remains invariant as long as the band gaps are preserved. Therefore, in the connection between the topologically nontrivial material and the trivial vacuum, there exist unavoidably the edge state modes (ESMs) spatially confining at the boundary and spectrally traversing the band gaps \cite{LaughlinPRB1981}. The presence of these gapless ESMs thus serves as an unambiguous signature of the topological non-triviality of the bulk band structure.

Recently, the concept of topology has been extended to circuit quantum electrodynamics (QED) lattice \cite{KochReview1,KochReview2}, where the electrons are replaced by microwave photons hopping between superconducting transmissionline resonators (TLRs) \cite{JQYouReview,DevoretReview2013}. While the idea of topological photonics was firstly developed in photonic crystals \cite{HaldaneTPPRL,TPNature2009,FanSHDMNP2012,HafeziNatPhoton2013,LuTPReviewNatPhoton2014}, circuit QED enjoys the time-resolved engineering of a large-scale lattice at the single-site level \cite{UCSBShellGameNP2011,UnderwoodDisorderPRA2012}. The demonstrated strong coupling between superconducting qubits and TLRs \cite{JQYouReview,DevoretReview2013} further allows the effective photon-photon interaction \cite{HartmanCoupledCavityNP2006,JCHNP2006,RebicKerr2009PRL,LangPhotonBlocakde2011PRL} which can hardly be achieved in other physical systems, indicating prospective future of investigating strongly correlated photonic liquids \cite{CarusottoQFLRMP2013,NoriQSReviewRMP2014}. As photons are charge neutral, there have been several proposals of synthesizing artificial magnetic fields on a TLR lattice, with predicted strengths much stronger than those in conventional electronic materials \cite{BoseFQHEPRL2008,YangWLGaugeFieldPRA2012,KochTRSPRA2010,PetrescuAQHEPRA2012}. Nevertheless, the synthetic Abelian gauge field has not been implemented so far despite the extensive theoretical studies, partially due to the complicated circuit elements required in these schemes. In addition, the detection of the integer topological invariants has also been addressed in recent research \cite{OzawaPRL2014,HafeziPRL2014,MeiFengPRA2015}, which is nontrivial in the sense that the Hall conductance measurement cannot be transferred to circuit QED due to the absence of fermionic statistics.

Here, we propose a theoretical scheme of implementing topological photonics in a two-dimensional circuit QED lattice. The distinct merit of our proposal is that we couple the TLRs by parametric frequency conversion (PFC) method which is simple in experimental setup and feasible with state-of-the-art technology \cite{DCEexperimentNature2011,NISTParametricConversionNP2011,NISTHongOuMandelPRL2012,NISTCoherentStateAPL2015}. The lattice in our scheme is formed by TLRs connected to the ground through the superconducting interference devices (SQUIDs) \cite{DCEexperimentNature2011,NISTParametricConversionNP2011,WangYPChiral2015,FelicettiPRL2014}, where the tunable photon hopping with nontrivial phases between TLRs can be induced through the dynamic modulation of the SQUIDs, allowing the arbitrary synthesization of time- and site-resolved gauge fields on the square lattice \cite{FanSHDMNP2012,WangYPChiral2015}. Moreover, with the driving-dissipation mechanism being employed, various quantities of the ESMs can be measured by the pumping and the steady-state photon number (SSPN) detection of only few sites on the lattice \cite{NISTCoherentStateAPL2015}.  In particular, the integer topological winding numbers of the ESMs with arbitrary values can be directly probed through the realization of the adiabatic pumping process \cite{LaughlinPRB1981,HafeziPRL2014}. Such measurement is equivalent to the measurement of the Chern numbers of the bulk bands and thus clearly examine the topological nontriviality of the photons. Furthermore, our detailed discussions show unambiguously that our proposal is very robust against various potential imperfection sources in experimental realizations due to the topological nature of the ESMs, pinpointing the feasibility with current level of technology.  Being flexible for the extension to more complicated lattice configurations and the incorporation of effective photon correlation, our scheme serves therefore as a promising and versatile platform for the future investigation of  various photonic quantum Hall effects.

\section*{Results}

\begin{figure}[tbh]
\begin{center}
    \includegraphics[width=0.7\textwidth]{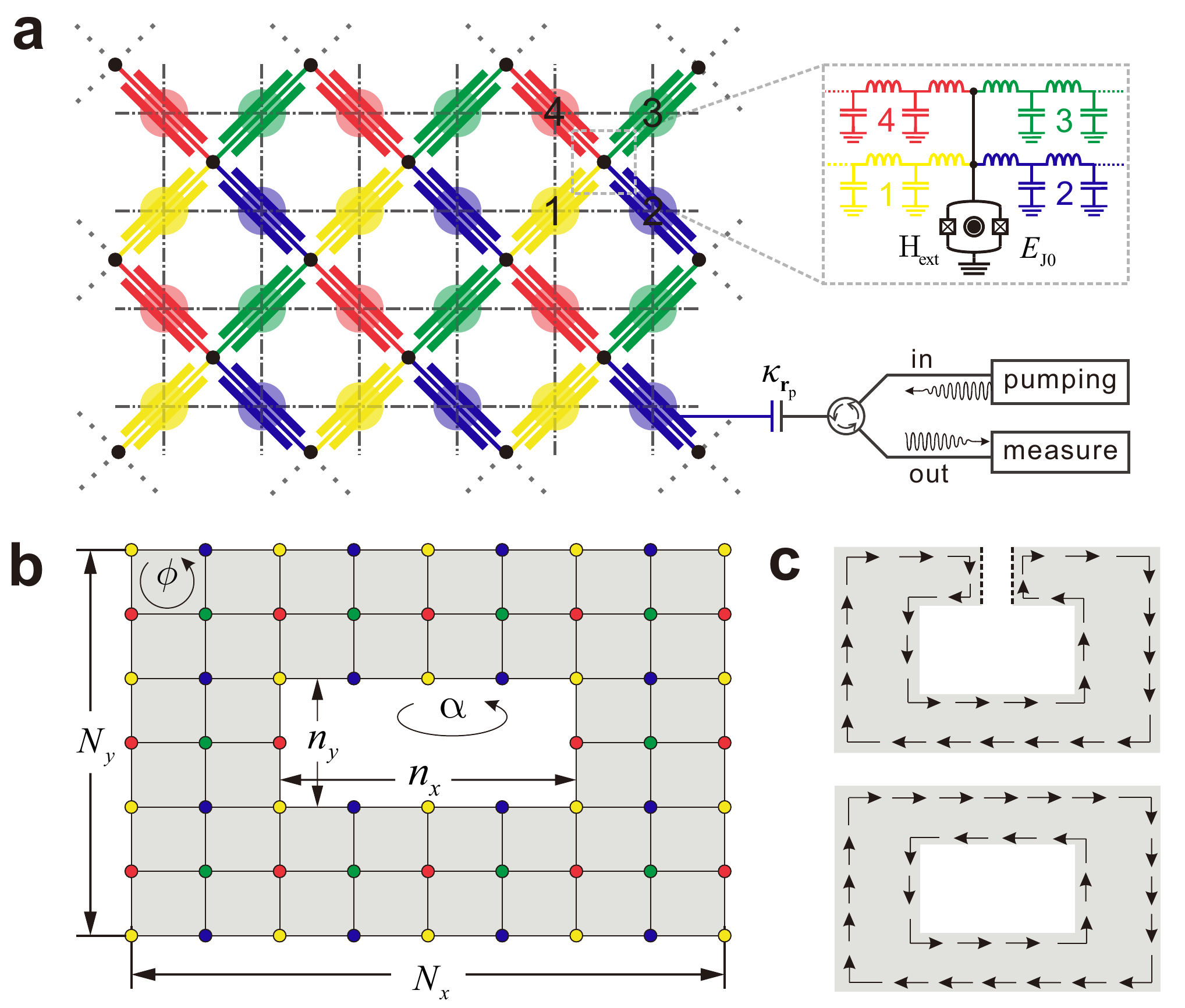}
\end{center}
\caption{\label{Fig Lattice} \textbf{(a)} Sketch of the square TLR lattice with the four colors (yellow, blue, green, and red) denoting the different lengths of the TLRs and the black dots representing the grounding SQUIDs (see the lumped circuit magnified representation at the right side). Each TLR plays the role of a photonic site (the large colored rounds) and the effective hopping between them (the dotted-dash lines) can be induced through the dynamic modulation of the SQUIDs. The pumping and the consequent steady-state measurement can be performed through the external coil connected to the pumping site(s) (lower right). \textbf{(b)} Configuration of the proposed lattice. The colored rounds and the solid lines label the TLRs with corresponding lengths and the photon hopping branches, respectively. \textbf{(c)} Spatial geometry of the lattice. The lattice shown in \textbf{(b)} can be obtained from the gluing of a simply-connected plane by the two dashed sides (the upper panel). Through this process the opposite chirals of the inner and the outer ESMs (the arrows) are formed (the lower panel).}
\end{figure}

\textbf{The lattice.}---We start with a square lattice consisting of TLRs with four different lengths placed in an interlaced form, as shown in Fig.~\ref{Fig Lattice}(a). At their ends, the TLRs are commonly grounded by SQUIDs with effective inductances much smaller than those of the TLRs \cite{FelicettiPRL2014,WangYPChiral2015,SupplementaryInformation,DCETheoryPRL2009}. Due to their very small inductances, the grounding SQUIDs impose low-voltage shortcuts at the ends of the TLRs. Therefore the lowest eigenmodes of the lattice can be approximated by the $\lambda/2$ modes of the TLRs with their ends being the nodes, and the whole lattice can be described by
\begin{equation}
\mathcal{H}_{\mathrm{S}}= \sum_{\mathbf{r}} \omega_{\mathbf{r}} a_{\mathbf{r}}^{\dagger} a_{\mathbf{r}},
\label{Eqn Hami}
\end{equation}
with $a_{\mathbf{r}}^{\dagger}$/$a_{\mathbf{r}}$ being the creation/annilhilation operators of the $\mathbf{r}$th photonic mode and $\omega_{\mathbf{r}}$ being the eigenfrequency. We further specify the eigenfrequencies of the four kinds of TLRs as yellow--$\omega_{\mathrm {0}}$, blue--$\omega_{\mathrm{0}}+\Delta$, green--$\omega_{\mathrm{0}}+3\Delta$, and red--$\omega_{\mathrm{0}}+4\Delta$, respectively, with $\omega_{\mathrm{0}}/2 \pi \in [10,20]$ $\mathrm{GHz}$ and $\Delta/2\pi \in [1,3/2]$ $\mathrm{GHz}$. Such configuration is for the following application of the dynamic modulation method and can be achieved through the length selection of the TLRs in the millimeter range \cite{SupplementaryInformation,NISTParametricConversionNP2011,NISTHongOuMandelPRL2012,NISTCoherentStateAPL2015}.

We then consider how to implement on the TLR lattice the effective tight-binding Hamiltonian
\begin{equation}
\label{Eqn aim}
\mathcal{H}_{\mathrm{T}} = \mathcal{T} \sum_{\langle \mathbf{r,r^{\prime}} \rangle} a_{\mathbf{r^{\prime}}}^{\dagger} a_{\mathbf{r}} e^{-i \theta_{\mathbf{r^{\prime}r}}} + \mathrm{h.c.},
\end{equation}
in the rotating frame of $\mathcal{H}_{\mathrm{S}}$. Here $\mathcal{T}$ is the uniform hopping amplitude, and
\begin{equation}
\theta_{\mathbf{r^{\prime}\mathbf{r}}}=\int_{\mathbf{r}}^{\mathbf{r^{\prime}}} \mathbf{A(x)}\cdot\mathrm{d}\mathbf{x},
\end{equation}
is the $\mathbf{r} \rightarrow \mathbf{r^{\prime}}$ hopping phase manifesting the presence of a vector potential $\mathbf{A(x)}$ through Peierls substitution \cite{BernevigBook}. For each plaquette of the lattice, the summation of the hopping phases around its loop has the physical meaning
\begin{equation}
\theta_{\mathrm{plaquette}}=\oint \mathbf{A(x)}\cdot\mathrm{d}\mathbf{x}=\iint \mathbf{B(x)}\cdot\mathrm{d}\mathbf{S},
\end{equation}
i. e. the synthetic local magnetic field for the microwave photons.

However, it is nontrivial to have complex hopping constants between TLRs because the physical coupling between two TLRs takes real coupling constants, no matter capacitive \cite{UnderwoodDisorderPRA2012,UCSBShellGameNP2011} or inductive \cite{NISTHongOuMandelPRL2012,NISTCoherentStateAPL2015}. We then consider the dynamic modulation method studied in recent experiments \cite{NISTParametricConversionNP2011,NISTHongOuMandelPRL2012,NISTHongOuMandelPRL2012,NISTCoherentStateAPL2015}. The grounding SQUIDs can be modeled as flux-tunable inductances and it is now experimentally possible to modulate the SQUIDs by a. c. magnetic flux oscillating at very high frequencies (the experiment-achieved range is typically $8 \sim 10$ GHz \cite{DCETheoryPRL2009,DCEexperimentNature2011} which is much higher than the following-proposed $1 \sim 6 $ GHz). Such a. c. modulation introduces a small a. c. coupling
 \begin{equation}
 \mathcal{H}_\mathrm{AC}=\sum_{\langle \mathbf{r} , \mathbf{r^{\prime}} \rangle} \mathcal{T}^{\mathrm{ac}}_{\mathbf{r} \mathbf{r^{\prime}}}(t)(a_{\mathbf{r}}+a_{\mathbf{r}}^{\dagger}) (a_{\mathbf{r^{\prime}}}+a_{\mathbf{r^{\prime}}}^{\dagger}),
  \end{equation}
 in addition to the d. c. contribution of the SQUIDs which  is irrelevant because the TLRs are largely detuned \cite{SupplementaryInformation}. We then assume that the a. c. modulation of the grounding SQUIDs contain three tones with frequencies being $\Delta$, $2\Delta$, and $4\Delta$. By bridging the frequency differences between the TLRs, the $2\Delta$/$4\Delta$ tones induce the vertical blue $\Leftrightarrow$ green/red $\Leftrightarrow$ yellow parametric hoppings, and the $\Delta$ tone establishes the horizontal yellow $\Leftrightarrow$ blue and green $\Leftrightarrow$ red parametric hoppings \cite{SupplementaryInformation}. When experiencing the PFC process between the TLR sites, the microwave photons adopt the phases of the a. c. modulating pulses, leading to the effective controllable complex hopping constants in the rotating frame of $\mathcal{H}_\mathrm{S}$ \cite{NISTCoherentStateAPL2015,WangYPChiral2015}. Through the application of the developed three-tone PFC pulses to each of the grounding SQUIDs, every vertical hopping branch and every pair of horizontal hopping branches can be independently controlled by a modulating tone threaded in one of the grounding SQUIDs \cite{SupplementaryInformation}, implying that the artificial magnetic field for microwave photons with Landau gauge
\begin{align}
\mathbf{A}=\left[ 0,A_y(x),0 \right], \mathbf{B}=B\mathbf{e}_z=\left[ 0,0,\frac{\partial}{\partial x} A_y(x) \right],
\end{align}
can be created with \textit{in situ} tunability. A further estimation demonstrates that the uniform hopping strength can be synthesized in the range $\left| \mathcal{T}/2\pi \right| \in \left[ 5, 15 \right] \text{ MHz} $ \cite{NISTParametricConversionNP2011,NISTHongOuMandelPRL2012,SupplementaryInformation}.

\begin{figure}[tbh]
\begin{center}
   \includegraphics[width=0.7\textwidth]{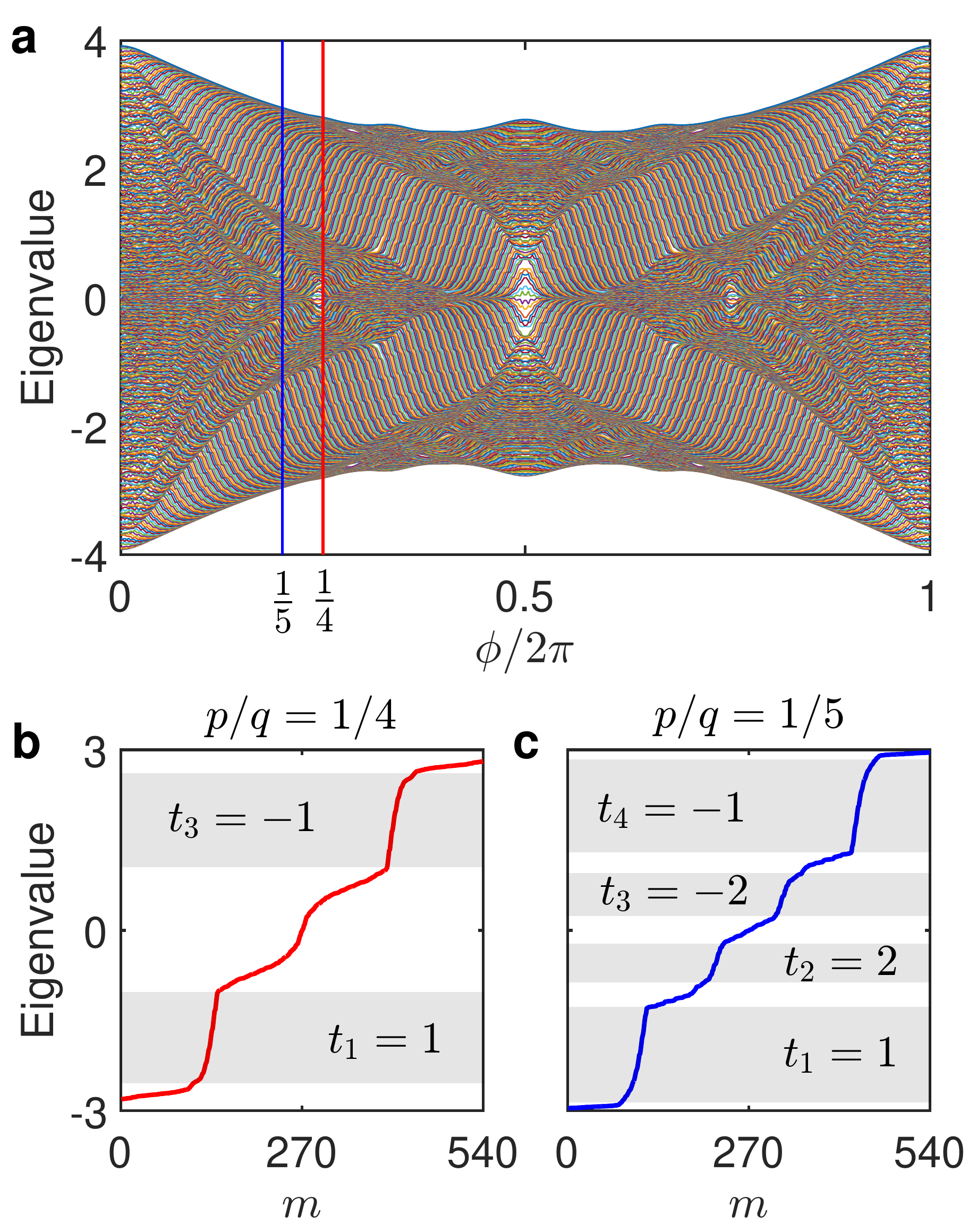}
\end{center}
\caption{\label{Fig Spectrum} \textbf{(a)} Hofstadter butterfly spectrum of the proposed lattice with $N_x\times N_y=24\times24$, $n_x\times n_y=6\times6$, and $\alpha/2\pi=0$, where energy is in units of $\mathcal{T}$. The situations of the rational effective magnetic fields $\phi/2\pi=1/4$ and $\phi/2\pi=1/5$ are denoted by the red and blue lines, respectively. Their eigenenergies are shown in \textbf{(b)} and \textbf{(c)}, with $m$ the index labeling the $540$ eigenvalues from smallest to largest. The band gaps are highlighted with their topological winding numbers marked.}
\end{figure}

For the investigation convenience of the ESM physics, in what follows we endow a nontrivial ring geometry to the TLR lattice, i.~e. an $N_{x} \times N_{y}$ square lattice with an $n_{x} \times n_{y}$ vacancy at its middle, as shown in Fig.~\ref{Fig Lattice}(b). Through the careful setting of the hopping phases, we penetrate a uniform effective magnetic flux $\phi$ in each plaquette of the lattice and an extra $\alpha$ at the central vacancy. In Fig.~\ref{Fig Spectrum}(a) the energy spectrum of a finite lattice is calculated with $N_{x}=N_{y}=24$ and $n_{x}=n_{y}=6$. In the rational situation $\phi/2\pi=p/q$ with $p$, $q$ being co-prime integers, the unit-cell of the lattice is enlarged by $q$ times, leading to $q$ nearly flat magnetic bands and the fractal Hofstadter butterfly spectrum \cite{HofstadterButterflyPRB1976} (Fig.~\ref{Fig Spectrum}(a)). These $q$ magnetic bands have nontrivial topological band structures, and between the $q$ bands there exist ESMs traversing the $q-1$ band gaps \cite{LaughlinPRB1981,BernevigBook,LuTPReviewNatPhoton2014}. The lattice spectrums of the situations of interest $p/q=1/4$ and $p/q=1/5$ are shown in Figs.~\ref{Fig Spectrum}(b) and  \ref{Fig Spectrum}(c), respectively, where the flatness of the band steps and the stiffness of the connections between the steps imply the degeneracy of the Landau levels and the spectral location of the ESMs.

\textbf{Probing the ESMs: The spatial and spectral information.}---Compared with fermionic electronic systems, the photonic nature of circuit QED allows multiple occupation of a particular mode at the same time and the non-equilibrium driving-dissipation competition. Here we propose the following scheme of probing the topological nature of the ESMs. With the detailed modeling being discussed in \textbf{Methods}, we emphasize that the physics behind is that the exotic properties of the ESMs result in the novel steady states of the lattice, and the information of the ESMs can be extracted from the SSPNs of only few sites on the lattice versus the pumping frequency and the pumping sites.

\begin{figure}[tbh]
\begin{center}
\includegraphics[width=0.96\textwidth]{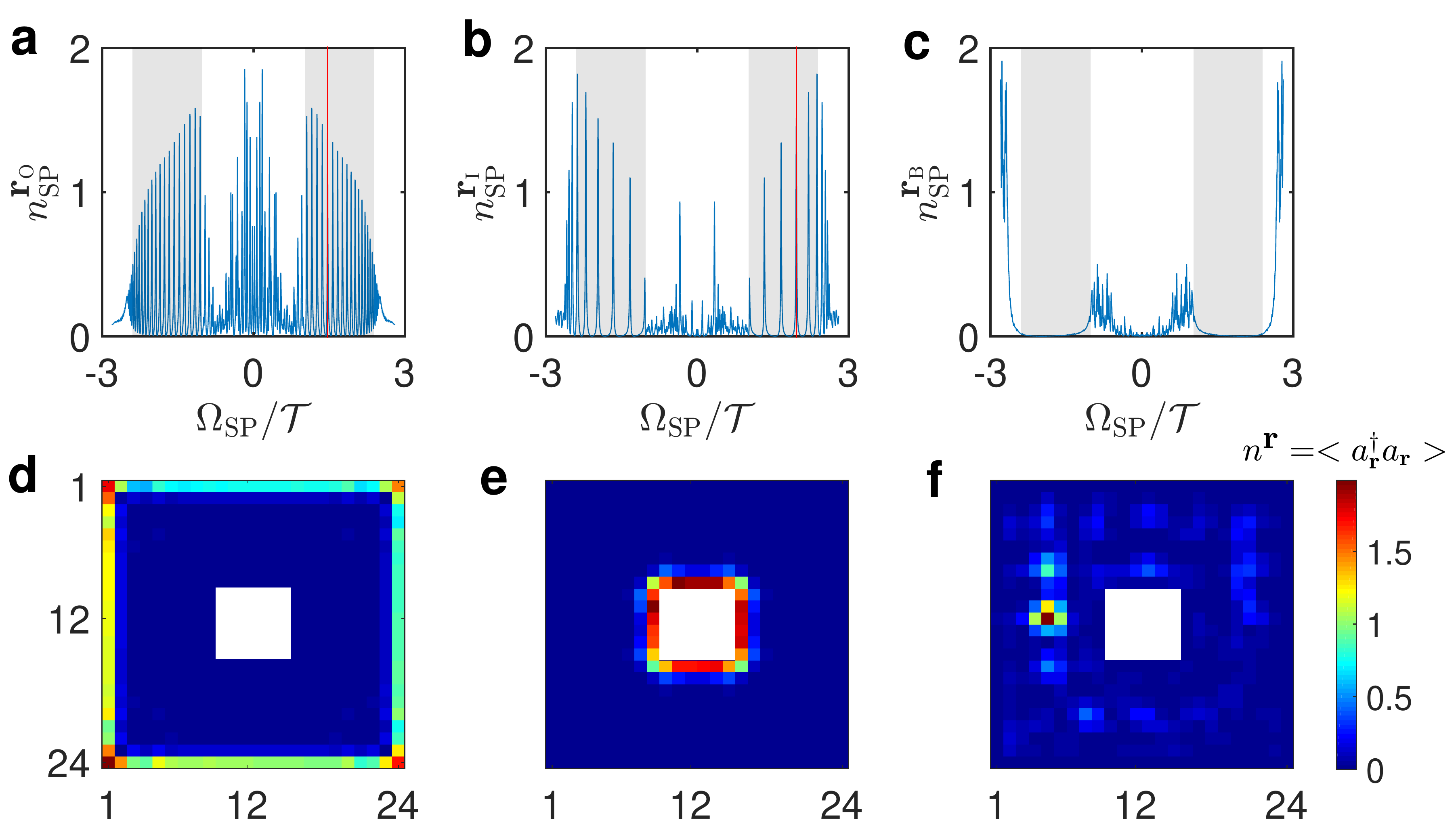}
\end{center}
\caption{\label{Fig Steady} Steady state of the proposed TLR lattice with $\phi/2\pi=1/4$, $\alpha=0$, and $\mathcal{T}/2\pi=10$ MHz. The SSPN on the pumping site $n_{\mathrm{SP}}^{\mathbf{r}_\mathrm{p}}$ versus $\Omega_{\mathrm{SP}}$  are displayed in \textbf{(a)}, \textbf{(b)}, and \textbf{(c)}, with $\mathbf{r}_\mathrm{p}=[(1,24)$, $(9,13)$, $(5,13)]$ and $\mathcal{P}_{\mathrm{S}}/\mathcal{T}=[0.5$, $0.25$, $0.23]$, respectively. The representative SSPN distributions on the whole lattice are presented in \textbf{(d)}---\textbf{(f)}, with the pumping frequencies $\Omega_{\mathrm{SP}}/\mathcal{T}=[1.47$, $1.97$, $2.69]$ marked by the corresponding red lines in \textbf{(a)}---\textbf{(c)}.}
\end{figure}

Firstly let us consider the single-site driving of a particular site $\mathbf{r}_{\mathrm{p}}$ described by
\begin{equation}
\label{Eqn SinglePump}
\mathcal{H}_{\mathrm{SP}}=\mathcal{P}_{\mathrm{S}}a_{\mathbf{r}_{\mathrm{p}}}e^{i\Omega_{\mathrm{SP}} t}+\mathrm{h.c.},
\end{equation}
with $\mathcal{P}_{\mathrm{S}}$ being the pumping strength and $\Omega_{\mathrm{SP}}$ being the detuning in the rotating frame of $\mathcal{H}_{\mathrm{S}}$. The SSPN on the pumping site $n_{\mathrm{SP}}^{\mathbf{r}_{\mathrm{p}}}=\langle a^{\dagger}_{\mathbf{r}_{\mathrm{p}}}a_{\mathbf{r}_{\mathrm{p}}}\rangle$ in the situation $p/q=1/4$ and $\alpha=0$ is numerically simulated based on equation (\ref{Eqn SSnumber}) and plotted in Fig. \ref{Fig Steady}. In what follows we show that the spatial and spectral information of the ESMs can be distilled by measuring the dependence of the single-site SSPN $\langle a^{\dagger}_{\mathbf{r}_{\mathrm{p}}}a_{\mathbf{r}_{\mathrm{p}}}\rangle$ on $\Omega_{\mathrm{SP}}$ and $\mathbf{r}_{\mathrm{p}}$.

If we choose $\mathbf{r}_{\mathrm{p}}=\mathbf{r}_{\mathrm{O}}=(1,24)$ as an outer edge site (OES), significant $n_{\mathrm{SP}}^{\mathbf{r}_{\mathrm{O}}}$ can be detected when $\Omega_{\mathrm{SP}}$ falls in the $1$st and $3$rd gaps, indicated by the highlighted spectrum comb in Fig.~\ref{Fig Steady}(a) (for the even $q=4$, the $2$nd gap is closed as a Dirac point form). This can be attributed to the excitation of the outer ESMs. However, if $\Omega_{\mathrm{SP}}$ is chosen deeply in the magnetic bands, $n_{\mathrm{SP}}^{\mathbf{r}_{\mathrm{O}}}$ has bare value because in this situation $\mathcal{H}_{\mathrm{SP}}$ can only excite bulk state modes (BSMs) which spread over the whole lattice, i.~e. the weight of $\mathbf{r}_{\mathrm{p}}$ in the mode function becomes diluted. The situation of pumping an inner edge site (IES) $\mathbf{r}_{\mathrm{p}}=\mathbf{r}_{\mathrm{I}}=(9,13)$ is similar, where the comb-like spectrum of $n_{\mathrm{SP}}^{\mathbf{r}_{\mathrm{I}}}$ centralized in the band gaps can also be found in Fig.~\ref{Fig Steady}(b). Meanwhile, there are still several interesting differences. As the number of the IESs is smaller than that of the OESs, Fig.~\ref{Fig Steady}(b) contains fewer peaks than Fig.~\ref{Fig Steady}(a) \cite{BernevigBook}. In addition, different pumping strengths have been used in the numerical simulation of  Figs.~\ref{Fig Steady}(a) and \ref{Fig Steady}(b) such that the obtained $n_{\mathrm{SP}}^{\mathbf{r}_{\mathrm{O}}}$ and $n_{\mathrm{SP}}^{\mathbf{r}_{\mathrm{I}}}$ are in the same region. This choice can also be traced back to the small number of the IESs which results in the concentration of the mode functions in the inner edge. Another observation is the opposite trends of $n_{\mathrm{SP}}^{\mathbf{r}_{\mathrm{O}}}$ and $n_{\mathrm{SP}}^{\mathbf{r}_{\mathrm{I}}}$ versus $\Omega_{\mathrm{SP}}$: In the $1$st  gap of Figs.~\ref{Fig Steady}(a) and (b), the peaks of $n_{\mathrm{SP}}^{\mathbf{r}_{\mathrm{O}}}$/$n_{\mathrm{SP}}^{\mathbf{r}_{\mathrm{I}}}$ increase/decrease with increasing $\Omega_{\mathrm{SP}}$. In contrast, when we set $\mathbf{r}_{\mathrm{p}}=\mathbf{r}_{\mathrm{B}}=(5,13)$ as a bulk site (BS), the lattice will have detectable $n_{\mathrm{SP}}^{\mathbf{r}_{\mathrm{B}}}$ iff $\Omega_{\mathrm{SP}}$ falls in the magnetic bands. When we choose $\Omega_{\mathrm{SP}}$ in the band gaps, the lattice cannot be excited because no BSM spectrally populates in the band gaps and no ESM spatially populates in the bulk of the lattice (notice the marked window at the $1$st and $3$rd gaps in Fig.~\ref{Fig Steady}(c)).

The above illustration can be experimentally detected by the proposed measurement scheme sketched in Fig. \ref{Fig Lattice}(a): A particular pumping site $\mathbf{r}_{\mathrm{p}}$ is capacitively connected to an external coil with input/output ports for pumping/measurement. The steady state of the lattice can be prepared by injecting microwave pulses through the input port for a sufficiently long time. During the steady-state period, energy will leak out of the $\mathbf{r}_\mathrm{p}$th TLR from the coupling capacitance, which is proportional $\omega_{\mathbf{r}_\mathrm{p}}\langle a_{\mathbf{r}_{\mathrm{p}}}^{\dagger}a_{\mathbf{r}_{\mathrm{p}}}\rangle$ with the proportional constant being determined by the coupling capacitance. The target observable $\langle a_{\mathbf{r}_{\mathrm{p}}}^{\dagger}a_{\mathbf{r}_{\mathrm{p}}}\rangle$ can therefore be measured by simply integrating the energy flowing to the output port in a given time duration. Actually, this measurement scheme has already been used in a recent experiment in which both the amplitude and the phase of a coherent state of a TLR were measured \cite{NISTCoherentStateAPL2015}. Here we emphasize that what we want to measure is the expectation value $\langle a_{\mathbf{r}_{\mathrm{p}}}^{\dagger}a_{\mathbf{r}_{\mathrm{p}}}\rangle$, while the detailed probability of the multi-mode coherent steady state projected to the Fock basis is nevertheless not needed. It is this weak requirement that greatly simplify our measurement.

We further calculate for each pumping situation a typical steady state photon distribution and display them in Figs.~\ref{Fig Steady}(d)--(f), respectively. In Figs. \ref{Fig Steady}(d) and \ref{Fig Steady}(e), the steady states correspond to the excitation of an inner or out ESM. The confinement and uniformity of the steady states clearly reflects the ESM mode functions localized and uniformly distributed on the edge, while the extended spatial distribution of the BS pumping steady state in Fig.~\ref{Fig Steady}(f) illustrates intuitively the difference between the BSMs and the ESMs.

\textbf{Probing the ESMs: Measuring the momentum.}---While the single-site pumping provides a route of discovering the spatial and spectral properties of the ESMs, the more interesting physics comes from the multi-site inhomogeneous pumping, which proves to be an efficient method of measuring the momentums of the ESMs. We pump $m$ consecutive OESs as
\begin{equation}
\label{Eqn Multipump}
\mathcal{H}_{\mathrm{MP}}=\mathcal{P}_{\mathrm{M}} \sum_{j=1}^{m} [a_{\mathrm{r}_{j}}e^{i(\Omega_{\mathrm{MP}}t-jk_{\mathrm{P}})}+\mathrm{h.c.}],
\end{equation}
and investigate the summed SSPN $n_{\mathrm{MP}}=\sum_{j=1}^{m} \langle a_{\mathrm{r}_{j}}^{\dagger}a_{\mathrm{r}_{j}} \rangle$ on the $m$ pumping sites. Here $\mathcal{P}_{\mathrm{M}}$ is the homogeneous pumping strength, $k_{\mathrm{P}}$ is the phase gradient of the pumping between neighboring sites, $\Omega_{\mathrm{MP}}$ is the frequency detuning in the rotating frame, and $\mathrm{r}_{j}$ for $j=1,2\ldots m$ denotes the $j$th of the $m$ pumping sites.

\begin{figure}[tbh]
\begin{center}
\includegraphics[width=0.7\textwidth]{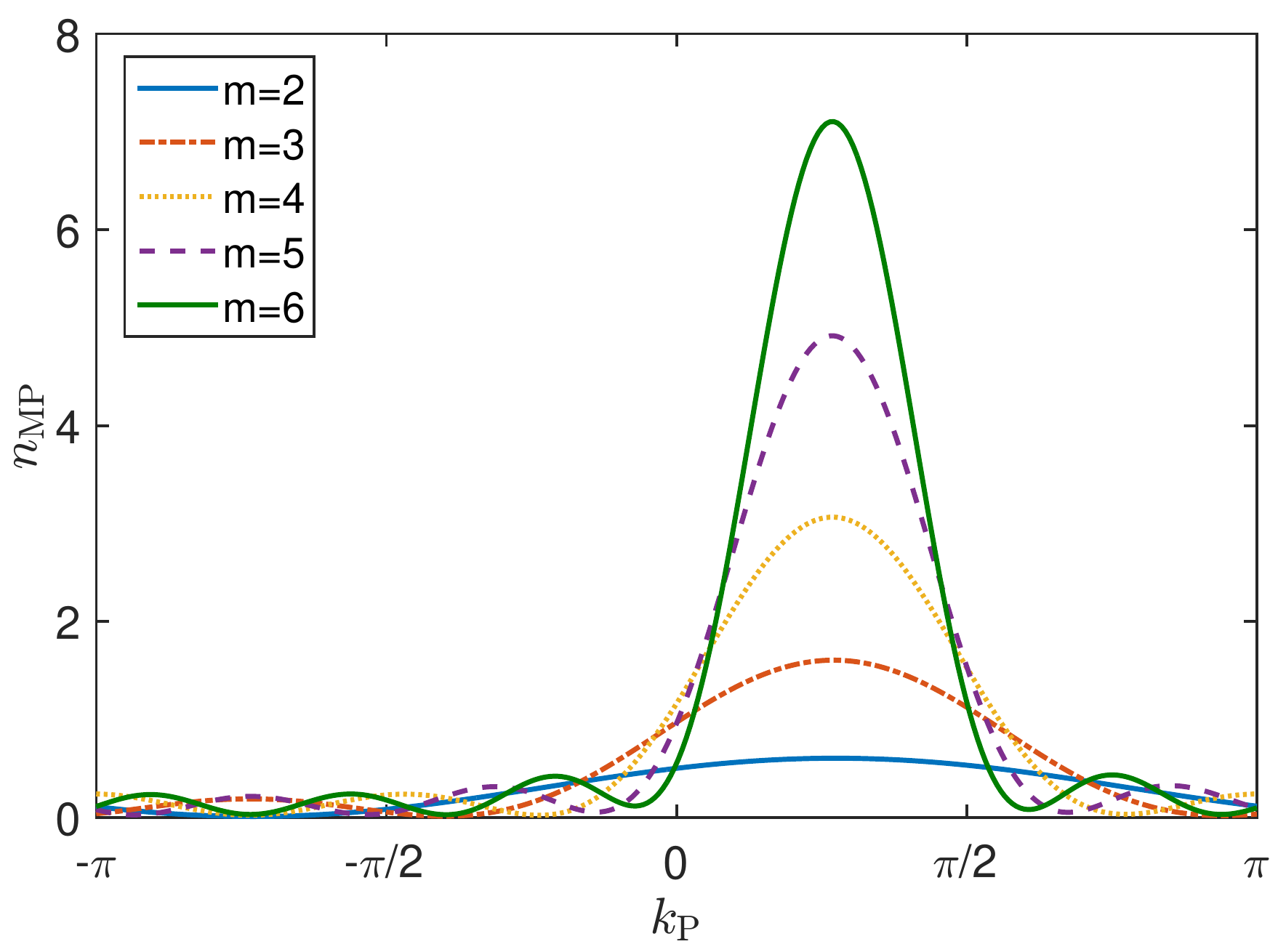}
\end{center}
\caption{\label{Fig Momentum} $n_{\mathrm{MP}}$ versus $k_{\mathrm{P}}$ and $m$ with $\mathcal{P}_{\mathrm{M}}/\mathcal{T}=0.1$ and $\Omega_{\mathrm{MP}}/\mathcal{T}=1.67$. The pumping sites start from the OES site $(\lfloor 7-m/2 \rfloor,24)$ and end at $(\lfloor 6+m/2 \rfloor,24)$. The other parameters are the same as those in Fig.~\ref{Fig Steady}.}
\end{figure}

Suppose $\Omega_{\mathrm{MP}}$ matches the eigenfrequency of a particular outer ESM, there arises an interesting question that, how does $n_{\mathrm{MP}}$ depend on $k_{\mathrm{P}}$? For a photon in that ESM, we can imagine its propagation around the edge with its ESM momentum $k_{\mathrm{0}}$ (this can be verified by the discussion of the coherent dynamics in \textbf{Discussion}). Therefore, the reduced ESM mode function on the $m$ pumping sites can be represented by a vector
\begin{equation}
\mathbf{k}_{\mathrm{0}}=(1,e^{ik_{\mathrm{0}}},e^{i2k_{\mathrm{0}}} \ldots e^{i(m-1)k_{\mathrm{0}}})
\end{equation}
where the $e^{ik_{\mathrm{0}}}$ factor denotes the phase delay between two consecutive sites and the equal-weighting character of $\mathbf{k}_{\mathrm{0}}$ reflects the uniform spatial distribution of the ESM on the confined edge (see Figs. \ref{Fig Steady}(d) and \ref{Fig Steady}(e)). It is this form of $\mathbf{k}_{\mathrm{0}}$ that inspires the inhomogenous multi-site pumping $\mathcal{H}_{\mathrm{MP}}$, which can be represented by another vector
\begin{equation}
\mathbf{k}_{\mathrm{P}}=(1,e^{ik_{\mathrm{P}}},e^{2ik_{\mathrm{p}}} \ldots e^{(m-1)ik_{\mathrm{p}}}).
\end{equation}
Based on the above physical picture, we can conjecture that the maximum of $n_{\mathrm{MP}}$ will emerge at the point
\begin{equation}
k_{\mathrm{P}}=k_{\mathrm{0}},
\end{equation}
where the excitations from the $m$ pumping sites constructively interfere with each other. The dependence of $n_{\mathrm{MP}}$ on $k_{\mathrm{P}}$ and $m$ is plotted in Fig.~\ref{Fig Momentum}, where the positions of the peaks infer the value of $k_{\mathrm{0}}$. In addition, the full width of half maximum (FWHM) of the peaks decreases with the increase of $m$. This can be understood by considering the two extreme cases: If $m=1$, there is certainly no peak because the steady state is independent on $k_{\mathrm{P}}$. Meanwhile, when all the OESs participate in the inhomogeneous pumping and the lattice size grows up, the inner product ${\vert \mathbf{k}_{\mathrm{P}}^{\dagger}\mathbf{k}_{\mathrm{0}}\vert}^{2}$ describing the interference between the pumping sites becomes nonzero iff $k_{\mathrm{P}}=k_{\mathrm{0}}$. In this situation the peaks in Fig.~\ref{Fig Momentum} approach a $\delta$-like function. As implied in Fig.~\ref{Fig Momentum}, for a moderate $m=5$ the FWHM is already sharp enough to discriminate the ESM momentums with a satisfactory resolution.

\textbf{Probing the ESMs: The integer topological invariants.}---We further consider the measurement of the integer topological quantum numbers of the system. The topological property of a electronic Bloch band is captured by the quantized Hall conductance which turns out to be its Chern number \cite{BernevigBook}. This transport measurement is nevertheless inaccessible in circuit QED systems due to the absence of Fermi statistics. Meanwhile, the presence of the ESMs provides an alternative way of probing the topological invariants according to the bulk-edge correspondence \cite{LaughlinPRB1981}. In the rational situation $\phi/2\pi=p/q$, the eigenenergies of the ESMs can be represented by the zero points of the Bloch functions winding around the $q-1$ holes of a complex energy surface which correspond to the $q-1$ gaps of the lattice \cite{BernevigBook}. The topological quantum numbers of the ESMs are given by the Diophantine equation
\begin{equation}
\label{Eqn diophantine}
h=s_{h}q+t_{h}p,|t_{h}| \leq q/2 ,
\end{equation}
where $h$ is the gap index and $t_{h},s_{h}$ are integers. Especially, $t_{h}$ is the topological winding number of the $h$th gap which is related to the Chern number $\mathcal{C}_{h}$ of the $h$th band as
\begin{equation}
\label{Eqn ChernNumber}
\mathcal{C}_{h}=t_{h}-t_{h-1}.
\end{equation}
Measuring the winding numbers of the ESMs is thus equivalent to measuring the Chern numbers of the magnetic bands. For $q=4$ we have $\mathcal{C}_{1}=\mathcal{C}_{3}=1$ and $t_{1}=-t_{3}=1$, while for $q=5$ we have $\mathcal{C}_{3}=-4$ and  $\mathcal{C}_{j}=1$ for the other four bands, and $t_{1}=-t_{4}=1$,$t_{2}=-t_{3}=2$ (see Figs. \ref{Fig Spectrum}(b) and \ref{Fig Spectrum}(c)).

\begin{figure*}[tbh!]
\begin{center}
\includegraphics[width=0.96\textwidth]{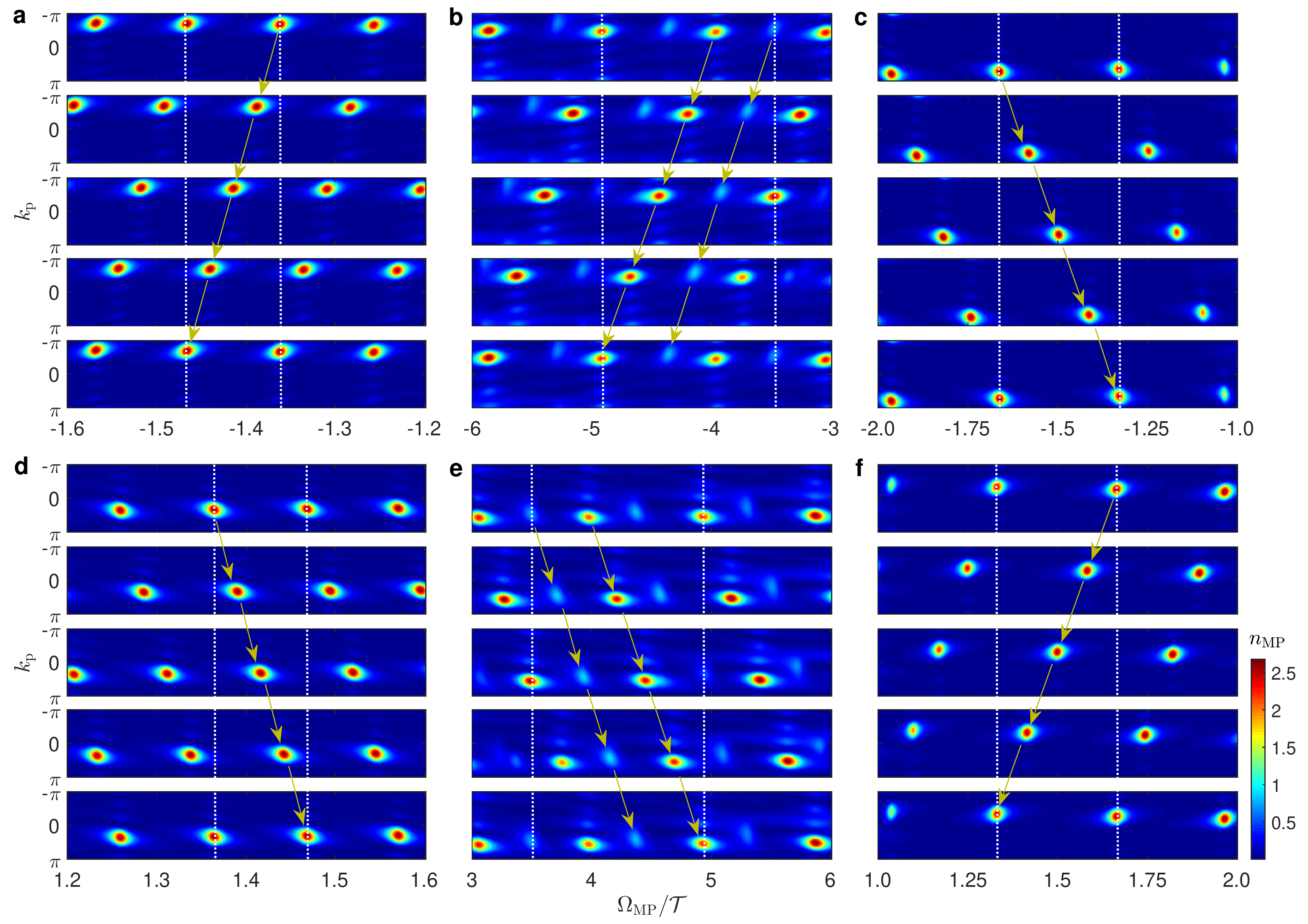}
\end{center}
\caption{\label{Fig Windingnumber} Adiabatic pumping of the proposed lattice represented by $n_{\mathrm{MP}}$ versus $\Omega_{\mathrm{MP}}$ and $k_{\mathrm{P}}$. Here we set $\phi/2\pi=1/4$ for \textbf{(a)}, \textbf{(c)}, \textbf{(d)}, and \textbf{(f)}, and $\phi/2\pi=1/5$ for \textbf{(b)} and \textbf{(e)}. The panels from top to bottom in each of the subfigures correspond to $\alpha/2\pi=0$, $1/4$, $1/2$, $3/4$ and $1$, respectively. For \textbf{(a)}, \textbf{(b)}, \textbf{(d)}, and \textbf{(e)} the OESs $(4,1)$---$(8,1)$ are pumped, while for \textbf{(c)} and \textbf{(f)} the IESs $(9,9)$---$(13,9)$ are pumped. Notice the ranges of the pumping frequency are chosen in a mirror form for the upper and lower subfigures. The other parameters are the same as those in Figs.~\ref{Fig Steady} and \ref{Fig Momentum}.}
\end{figure*}

As the spatial configuration of the proposed lattice is equivalent to the Laughlin cylinder \cite{LaughlinPRB1981,BernevigBook} (i. e. it can be regarded as the rolling of a two-dimensional simply-connected plane and the consequent threading with an effective magnetic flux $\alpha$, see Fig.~\ref{Fig Lattice}(c)), the adiabatic pumping of the ESMs can be realized through the control of $\alpha$ in the central vacancy. Once $\alpha$ increases monotonically from $0$ to $2\pi$, an integer number of ESMs will be transferred with the spectrum of the lattice returned to its original form. Such integer is exactly the winding number of the ESMs. In Figs.~\ref{Fig Windingnumber}(a) and \ref{Fig Windingnumber}(d) the dependence of $n_{\mathrm{MP}}$ on $k_{\mathrm{P}}$ and $\alpha$ is numerically calculated for $p/q=1/4$. Guided by the dashed white lines and the solid yellow arrows, the ESM peaks in the $1$st and $3$rd gaps move by one with the opposite moving directions, in agreement with the relation $t_{1}=-t_{3}=1$. This scheme is in principle general to measure integer topological invariants with any value: The peaks of the ESMs in the $h$th gap move by $|t_{h}|$ during the variation of $\alpha$, with the moving direction indicating the sign of $t_{h}$, and the Chern numbers $\mathcal{C}_{h}$ can be calculated from equation~(\ref{Eqn ChernNumber}) after all $t_{h}$ are obtained. The situation $\phi/2\pi=1/5$ is also shown in Figs.~\ref{Fig Windingnumber}(b) and \ref{Fig Windingnumber}(e), where the movements in the $2$nd and $3$rd gaps cross two peaks, indicating $t_{2}=-t_{3}=2$. In addition, the opposite moving directions can also be observed in the pumping of the inner ESMs and the outer ESMs in the same gap (see Figs.~\ref{Fig Windingnumber}(c) and \ref{Fig Windingnumber}(f)).

\section*{Discussion}

\textbf{Robustness against imperfection factors.}---The imperfection in realistic experiments accompanies inescapably with the ideal scheme proposed above, including  the diagonal and off-diagonal disorders $\omega_{\mathbf{r}} \rightarrow \omega_{\mathbf{r}}+\delta\omega_{\mathbf{r}},\mathcal{T} \rightarrow \mathcal{T}+\delta\mathcal{T}_\mathbf{r^\prime r}$
added into $\mathcal{H}_{\mathrm{T}}$ by the fabrication errors of the circuit and the low-frequency $1/f$ background noises \cite{FlickerRMP2014}, and the non-nearest-neighbor hoppings  induced the residual long-range coupling between TLRs which are not presented in the proposed $\mathcal{H}_{\mathrm{T}}$. Understanding these effects is thus crucial for our scheme.  As briefly summarized below and detailedly studied in ref. \cite{SupplementaryInformation}, these imperfection factors lead to unwanted terms which are all much smaller than the band gaps of the lattice ($\sim \mathcal{T}$). Moreover, some of the resulted effects can be further suppressed by the slight refinement of the proposed scheme. Such small fluctuations cannot destroy the topological properties of the ESMs because they are not strong enough to close and reopen the band gaps. Therefore, the presence of these imperfections can only renormalize negligibly the predicted results. From this point of view, our scheme enjoys the topological protection against the imperfection factors, which pinpoints its feasibility based on current level of technology.

In particular, the fabrication error including the deviations of the realized circuit parameters from the ideal settings (e. g. the lengths and the unit capacitances or inductances of the TLRs) leads to the disorder of the eigenmodes' frequencies $\delta\omega_\mathbf{r}$, while the long-range coupling induced by the finite inductances of the grounding SQUIDs results in the next-nearest-neighbor hoppings through the current-division mechanism \cite{MartinisXmonPRL2014,MartinisXmonPRA2015}. As evaluated in ref. \cite{SupplementaryInformation}, these effects are both at the level of  $\left[10^{-2},10^{-1}\right]\mathcal{T}$. In addition, these errors can be further corrected by simple revision of the proposed PFC scheme, including the refined choice of the TLRs' lengths and the corresponding renormalization of the modulating frequencies of the grounding SQUIDs \cite{SupplementaryInformation}.

In actual experimental circuits, the $1/f$ noise at low frequencies far exceeds the thermodynamic noise \cite{FlickerRMP2014}. The $1/f$ noise in superconducting quantum circuits can generally be traced back to the fluctuations of three degrees of freedom, namely, the charge, the flux, and the critical current. Due to its low frequency property, we can treat the $1/f$ noises as quasi-static, i. e. the noises do not vary during a experimental run, but vary between different runs.  Firstly, the proposed circuit is insensitive to the charge noise as it consists of only TLRs which are linear elements and grounding SQUIDs which have very small charging energies and very large effective Josephson coupling energies. Such insensitivity roots in the same origin of the charge insensitivity of transmon qubits \cite{KochTransmonPRA2007} and the flux insensitivity of the low decoherence flux qubit \cite{JQYouFluxBitPRB2007}. Secondly, the flux $1/f$ noises penetrated in the loops of the grounding SQUIDs shift the d. c. bias points of the grounding SQUIDs in a quasi-static way. The consequent effect is then the fluctuations
\begin{align}
\delta \omega_{\mathbf{r}} < 10^{-3} \mathcal{T},
\delta \mathcal{T}_{\mathbf{r^\prime r}} < 10^{-4} \mathcal{T},
\end{align}
where the detailed evaluation can be found in ref. \cite{SupplementaryInformation}. These flux-noise-induced diagonal and off-diagonal fluctuations are both much smaller than the band gaps and the spectral spacing between the ESM peaks ($\sim 10^{-1} \mathcal{T}$, see Fig. \ref{Fig Windingnumber}). Such small fluctuations can thus neither destroy the topological properties of the ESM nor mix the resolution of the ESMs in the SSPN measurement.  Therefore, we come to the conclusion that our scheme can survive in the presence of the $1/f$ flux noise. The influence of the critical current noise is similarly analyzed, with results indicating that the induced effects are even smaller than those of the flux $1/f$ noises and can then be safely neglected \cite{MartinisFlickerPRL2007,SupplementaryInformation}.

\begin{figure}[tbh]
\begin{center}
\includegraphics[width=0.62\textwidth]{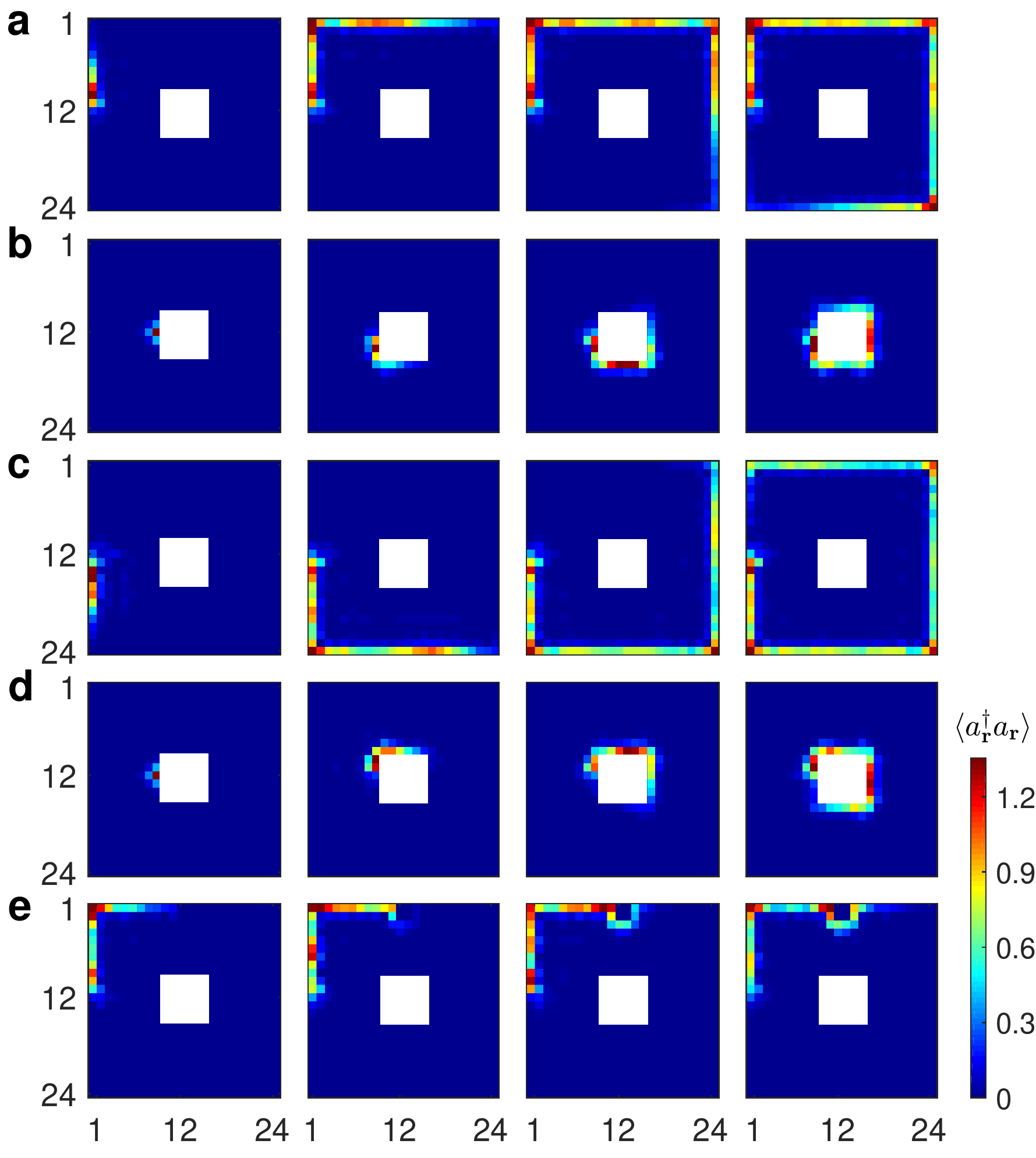}
\end{center}
\caption{\label{Fig Chiral} Chiral flow dynamics in the presence of single-site pumping. We set $\mathbf{r}=(1,13)$, $\mathcal{P}_{\mathrm{S}}/\mathcal{T}=2$ for \textbf{(a)}, \textbf{(c)}, and \textbf{(e)}, and $\mathbf{r}=(9,13)$, $\mathcal{P}_{\mathrm{S}}/\mathcal{T}=1.5$ for \textbf{(b)} and \textbf{(d)}. The synthesized magnetic field is set as $\phi/2\pi=1/4$ and $\alpha/2\pi=0$. In addition, $\Omega_{\mathrm{SP}}/\mathcal{T}$ is chosen as $[-1.76$, $-1.97$, $1.47$, $1.97$, $-1.75]$ for \textbf{(a)}--\textbf{(e)}, respectively. The times of the panels from left to right are arithmetic progressions with the first terms $T_1/2\pi=[6$, $1$, $6$, $1$, $15]\mathcal{T}^{-1}$ and the common differences $\Delta T/2\pi=[13$, $5$, $13$, $5$, $2.5]\mathcal{T}^{-1}$ for \textbf{(a)}--\textbf{(e)}, respectively. Especially, in the calculation of \textbf{(e)} the effect of lattice disorder and defect is incorporated (see the main text). The other parameters are set the same as those used in Fig.~\ref{Fig Steady}.}
\end{figure}

\textbf{Coherent chiral photon flow dynamics.}---A natural further step beyond the previous SSPN investigation is the study of coherent dynamics of the lattice, which is becoming experimentally possible due to the recent extension of the coherence times of superconducting circuits \cite{JQYouReview,DevoretReview2013}. Such investigation can offer an intuitive insight into the chiral property of the ESMs. We assume that the lattice is initially prepared in its ground state and then a driving $\mathcal{H}_{\mathrm{SP}}$ is added with $\mathbf{r}_\mathrm{p}$ being an edge state and $\Omega_{\mathrm{SP}}$ being the eigenfrequency of an ESM. The time evolution of the lattice is calculated with screenshots of the photon flow dynamics shown in Fig.~\ref{Fig Chiral}. The chirals of the ESMs result in the unidirectional photon flows around the edge with the directions determined by $\Omega_{\mathrm{SP}}$ and $\mathbf{r}_\mathrm{p}$. Chosen $\mathbf{r}_\mathrm{p}=\left(1,13\right)$ as an OES, the photon flow is clockwise/counterclockwise if $\Omega_{\mathrm{SP}}$ falls in the 1st/3rd gap (Figs.~\ref{Fig Chiral}(a) and \ref{Fig Chiral}(c)), indicating the opposite chirals of the ESMs in different gaps. The energy separation of ESMs with different chirals can be understood by regarding the unidirectional flow as a rotating spin \cite{HafeziNatPhoton2013}. Placed in an artificial magnetic field $\mathbf{B}$, such spin has split energies with one spinning direction lower and the opposite direction higher. Moreover, it is observed from Figs.~\ref{Fig Chiral}(a) and \ref{Fig Chiral}(b) (see also Figs.~\ref{Fig Chiral}(c) and \ref{Fig Chiral}(d)) that the chiral of the inner ESMs is opposite to that of the outer ESMs in the same gap. This oppositeness can be explained by the spatial configuration of the lattice. As shown in the upper panel of Fig.~\ref{Fig Lattice}(c), by ``tearing'' the lattice apart we can get a simply-connected plane where there is only one edge existing. Now we consider the inverse, i. e. we ``glue'' the two sides marked by dashed lines together, and obtain the ring geometry shown in Figs.~\ref{Fig Lattice}(a)  and the lower panel of \ref{Fig Lattice}(c). During this gluing, the ESM flow (marked by the arrows) cancels itself on the glued sides, leaving two closed circulations with opposite directions. This ``tearing-and-gluing'' process can also be tested in our lattice configuration because the horizontal hopping branches can be adiabatically tuned on and off. Another interesting observation is the different velocities of the photon flows between Figs.~\ref{Fig Chiral}(a) and \ref{Fig Chiral}(c). Notice the pumping frequencies in these two subfigures are not mirror to each other, such difference in flowing velocity reflects the momentum difference between the corresponding ESMs.

The chiral flow in the presence of disorder and defect is also calculated. Here we assume that $\delta\omega_\mathbf{r}$ and $\mathcal{T}_\mathbf{r^{\prime}r}$ are normally distributed with $\sigma(\delta\omega_\mathbf{r})=\sigma(\mathcal{T}_\mathbf{r^{\prime}r})=0.05\mathcal{T}$ much larger than those estimated in ref. \cite{SupplementaryInformation}. In addition, a $2 \times 2$ hindrance is placed on the upper outer edge with $\delta\omega_{(12-13,23-24)}/\mathcal{T}=30$. As displayed in Fig.~\ref{Fig Chiral}(e), the survival of the chiral photon flow under disorder and its circumvention around the hindrance clearly verify the topological robustness of the ESMs.

We remark that the proposed unidirectional photon flow can also be detected through the photon-number measurement of only few sites neighboring to the pumping sites or the defect sites: We first establish the PFC process and pump the lattice with $\mathcal{H}_\mathrm{SP}$ for a duration less than the time scale during which the photon flow circulates around the whole edge loop and then remove them.  The energy leaked out from the edge sites neighboring to the pumping sites can be observed, which is proportional to the photon numbers stored in those TLRs. From Fig. \ref{Fig Chiral} we expect that the injected photons tends to flow towards the sites on a particular direction with phase delay while leaves the sites on the opposite direction negligibly excited \cite{KochTRSPRA2010,WangYPChiral2015}. The comparison of the measured photon numbers on the opposite directions thus reveals the chiral property of the ESMs. The circumvention of the photon flow around the hindrance can be detected in a similar manner by measuring the sites neighboring to the hindrance sites.

\textbf{Extension in the future.}---The flexibility of the proposed PFC method is not limited by the square lattice configuration in this paper. For instance, a brick wall lattice (i. e. a stretched honeycomb lattice) can be straightforwardly obtained by closing some of the vertical hopping branches in the original square lattice. This generalization may pave an alternative way to the study of photonic graphene \cite{SegevPG2013}. In addition, while the cross-talk between the diagonal next-nearest-neighbor TLRs is suppressed in our scheme due to frequency mismatch, it can indeed be opened by adding another tone to the modulating pulses of the SQUIDs. This may offer potential facilities in the future study of anomalous quantum Hall effect in the checkerboard lattice configuration \cite{PetrescuAQHEPRA2012}.

In recent research, lattice configurations supporting a dispersionless flat band have been investigated extensively, including the Lieb and the Kagom\'e lattices \cite{PetrescuAQHEPRA2012}. These band structures provide an idea platform of achieving strongly-correlated phases as the kinetic energy is quenched \cite{BergholtzFB2013,SondhiFB2013}. These lattice configurations can also be synthesized by the variation of the proposed square lattice. While the photonic topological insulator considered in this paper can be understood in the single-particle picture, the introduction of interaction significantly complicates the problem and may lead to much richer but less explored physics. On the other hand, with the demonstrated strong coupling between TLRs and superconducting qubits \cite{JQYouReview,DevoretReview2013} (and also atomic system, see ref. \cite{NoriHybridReviewRMP2013}), the Bose-Hubbard \cite{HartmanCoupledCavityNP2006,RebicKerr2009PRL} and Jaynes-Cummings-Hubbard nonlinearities \cite{JCHNP2006,LangPhotonBlocakde2011PRL} can be incorporated into the proposed lattice. A further research direction should therefore be the implementation of photonic fractional Chern insulators and the understanding of strong correlation in the proposed architecture and its potential hybrid-system generalizations which may utilize the advantages of different physical systems \cite{NoriHybridReviewRMP2013,MaciejkoFTI2015}.

\section*{Conclusion}

In conclusion, we have proposed a method of implementing topological photonics in a circuit QED lattice. The effective magnetic field for microwave photons can be synthesized through the proposed parametric approach, and the topological properties of the ESMs can be extracted from the steady states of the lattice under pumping. Moreover, being flexible to incorporate effective photon-photon interaction, our proposal may offer a new route towards the investigation of nonequilibrium photonic quantum Hall fluids in on-chip superconducting quantum circuits. Taking the advantage of simplicity in setup and topological robustness against potential imperfections, the realization of this scheme is envisaged in the future experiments.

\section*{Methods}

\textbf{Steady state of the lattice.}---We explicitly consider the pumping of the lattice described by $\mathbf{P}^{\dagger}\mathbf{a}e^{-i\Omega t}+\mathrm{h.c.}$ in the rotating frame of $\mathcal{H}_{\mathrm{S}}$. Here $\mathbf{P}$ and $\mathbf{a}$ are the vectors composed of the pumping strengths and the annihilation operators of the lattice sites, respectively, and $\Omega$ is the monochromatic detuning playing the role of Fermi surface. In the presence of dissipation, the evolution of the lattice is described by the master equation
\begin{align}
\label{Eqn masterequation}
\frac{\mathrm{d}\rho }{\mathrm{d}t}&=-i\left[ \mathbf{a}^{\dagger}(\mathcal{B}-\Omega\mathcal{I})\mathbf{a}
+\mathbf{P}^{\dagger}\mathbf{a}+\mathbf{a}^{\dagger}\mathbf{P},\rho\right] \notag\\ &+\frac{1}{2}\sum_{\mathbf{r}} \kappa_{\mathbf{r}} \left(2a_\mathbf{r}\rho a_{\mathbf{r}}^{\dagger}-a_\mathbf{r}^{\dagger}a_\mathbf{r}\rho-\rho a_\mathbf{r}^{\dagger}a_\mathbf{r}\right),
\end{align}
where $\rho$ is the density matrix of the lattice, $\kappa_{\mathbf{r}}$ is the decay rate of the $\mathbf{r}$th TLR, and the matrix $\mathcal{B}$ is defined by $\mathbf{a}^{\dagger}\mathcal{B}\mathbf{a}=\mathcal{H}_{\mathrm{T}}$. As a linear system (i.~e. there is no photon-photon interaction), the lattice can be described in the picture of multi-mode coherent state, and its steady state can thus be determined by
\begin{equation}
\label{Eqn SSnumber}
i\frac{\mathrm{d}{\langle \mathbf{a}  \rangle}}{\mathrm{d}t}
=\left[\mathcal{B} -\Omega\mathcal{I}- \frac{1}{2} i\mathcal{K} \right] \langle \mathbf{a} \rangle + \mathbf{P}=0,
\end{equation}
where $\mathcal{K}$ is the diagonal matrix of the TLRs' decay rates. From equation (\ref{Eqn SSnumber}) our idea emerges that, $\Omega$ can be used to select the mode we are interested in, and the information of that mode can be extracted from the dependence of $\langle \mathbf{a} \rangle$ on $\mathbf{P}$. In addition, as the pumping sites are coupled to external coil, they suffer more severe decoherence than the other conventional sites. Therefore, we set the decay rates of the pumping sites as uniformly $\kappa_{\mathbf{r}_{\mathrm{p}}}/2\pi=2$ $\mathrm{MHz}$ and those of the conventional sites as uniformly $\kappa_{\mathbf{r}}/2\pi=100$ $\mathrm{kHz}$ throughout the numerical simulation of this paper, i. e. there is a $20$ times difference between them  \cite{NISTParametricConversionNP2011,NISTHongOuMandelPRL2012,NISTCoherentStateAPL2015}.

\begin{acknowledgments}
We thank Z. D. Wang@HKU, M. Gong@CUHK, Z. Q. Yin@THU, and L. Y. Sun@THU for helpful discussions. This work was supported in part by the National Fundamental Research Program of China (Grants No.~2012CB922103 and No.~2013CB921804), the National Science Foundation of China (Grants No.~11374117, No. 11574353, and No.~11375067), and the PCSIRT (Grant No.~IRT1243).
\end{acknowledgments}

\section*{Author Contributions}
Y.H and Z.Y.X proposed the idea. Y.P.W carried out all calculations under the guidance of Y.H. Z.Y.X, W.L.Y, and Y.W participated in the discussions. Y.H, Y.P.W and Z.Y.X contributed to the interpretation of the work and the writing of the manuscript.

\section*{Competing Interests}
The authors declare that they have no competing financial interests.


\clearpage

\clearpage

\clearpage

\clearpage

\clearpage

\clearpage

\end{document}